\documentclass[conference]{IEEEtran}
\IEEEoverridecommandlockouts
\usepackage{cite,amsmath,amssymb,amsfonts}
\usepackage{algorithmic,graphicx,textcomp,xcolor,tabularx,multirow,array}
\usepackage{caption}
\usepackage{subcaption}
\usepackage[utf8]{inputenc}
\usepackage[english]{babel}
\usepackage{url}
\renewcommand{\arraystretch}{1.5}

\def\BibTeX{{\rm B\kern-.05em{\sc i\kern-.025em b}\kern-.08em
    T\kern-.1667em\lower.7ex\hbox{E}\kern-.125emX}}
    
\begin{document}

\title{IoT Device Identification Based on Network Traffic Characteristics}

\author{\IEEEauthorblockN{Md Mainuddin, Zhenhai Duan}
\IEEEauthorblockA{Department of Computer Science\\
Florida State University\\
Tallahassee, FL 32306, USA\\
\textit{\{mainuddi, duan\}@cs.fsu.edu}%
}
\and
\IEEEauthorblockN{Yingfei Dong\\}
\IEEEauthorblockA{Department of Electrical Engineering\\
University of Hawaii\\
Honolulu, HI 96822 USA\\
\textit{yingfei@hawaii.edu}%
}
\and
\IEEEauthorblockN{Shaeke Salman, Tania Taami}
\IEEEauthorblockA{Department of Computer Science\\
Florida State University\\
Tallahassee, FL 32306, USA\\
\textit{\{salman, taami\}@cs.fsu.edu}%
}
}

\maketitle

\begin{abstract}
IoT device identification plays an important role in monitoring and improving the performance and security of IoT devices. Compared to traditional non-IoT devices, IoT devices provide us with both unique challenges and opportunities in detecting the types of IoT devices. Based on critical insights obtained in our previous work on understanding the network traffic characteristics of IoT devices, in this paper we develop an effective machine-learning based IoT device identification scheme, named {\em iotID}. In developing iotID, we extract $70$ features of TCP flows from three complementary aspects: remote network servers and port numbers, packet-level traffic characteristics such as packet inter-arrival times, and flow-level traffic characteristics such as flow duration. Different from existing work, we take into account the imbalance nature of network traffic generated by various devices in both the learning and evaluation phases of {\em iotID}. Our performance studies based on network traffic collected on a typical smart home environment consisting of both IoT and non-IoT devices show that iotID can achieve a balanced accuracy score of above $99\%$.
\end{abstract}


\section{Introduction}\label{sec:introduction}
Internet of Things (IoT) devices have been increasingly deployed in various environments, including smart homes, smart cities, and general enterprise networks. On the one hand, IoT devices greatly improve the convenience, efficiency, and quality of our lives and our communities; on the other hand, they also introduce many new challenges in monitoring and managing such devices, due to the immense diversity of manufacturers and models of IoT devices, and the sheer volume of such devices on the Internet. In this paper, we adopt a commonly accepted definition of IoT devices---an IoT device performs a specific functionality, and can operate autonomously without direct human controls. Examples of IoT devices include security cameras, smart plugs, and smart bulbs, to name a few. In contrast, traditional computers, laptops, and smart phones are examples of non-IoT devices; they can perform a diverse set of different functionalities directly instructed by the users of these devices.

One of the key foundations in monitoring and managing IoT devices is the capability to identify the specific type of an IoT device, including its manufacturer and model, termed {\em IoT device identification}. IoT device identification plays a critical role in monitoring and improving the performance and security of IoT devices. For example, a network may only allow certain types of IoT devices to be deployed inside the network; in addition, certain types of IoT devices can be blocked on a network once a security vulnerability is identified on such devices. Compared to traditional non-IoT devices, IoT devices provide us with both unique challenges and opportunities in detecting the types of IoT devices, including the vast diversity of IoT device types and mostly autonomous operations of specific (simple) functionalities. Given the importance of this problem, a number of IoT device identification techniques have been developed in the literature (see Section~\ref{sec:related-work} for details). However, these existing techniques have a few shortcomings in identifying IoT devices. 


First, most of these techniques considered networks comprising only IoT devices, without non-IoT devices. However, in a real-world deployment of an IoT device identification technique, a network normally consists of both IoT and non-IoT devices, and the identification technique must handle both IoT and non-IoT devices. Second, different IoT devices (and non-IoT devices) generate vastly different volumes of data traffic; put in another way, the data traffic of different devices is imbalanced. For example, security cameras in general generate much more traffic than smart plugs. However, none of the existing work considered the imbalance nature of datasets in either learning or evaluating the IoT device identification models. Third, the majority of the existing work developed machine-learning identification methods based on a large set of traffic features, for example, all the TCP/IP header fields. They failed to illustrate why and how a feature contributed to the overall performance of the developed identification method.

In this paper, we develop an effective machine-learning (ML) based IoT device identification approach, named {\em iotID}. We consider a realistic network environment consisting of both IoT and non-IoT devices, and iotID can effectively identify IoT devices based on the combined traffic of both kinds of devices. In addition, we also take into account the nature of imbalanced traffic generated by different kinds of devices in developing and evaluating iotID. Furthermore, the development of iotID is based on our previous work on studying the network traffic characteristics of IoT devices~\cite{mainuddin2021traffic}, which provides us with an informed understanding of the behavioral characteristics of different devices and critical insights in developing an effective IoT device identification scheme. We evaluate the performance of iotID using a week-long data trace collected in a typical smart home environment, comprising both IoT and non-IoT devices. Our evaluation studies show that iotID can achieve a balanced accuracy score of above $99\%$ in identifying the deployed IoT devices in the smart home environment.

The remainder of the paper is organized as follows. In Section~\ref{sec:problemstatement}, we describe the problem statement, and the collection and pre-processing of the data used in this study. In Section~\ref{sec:iotid} we present the design of the iotID scheme. We evaluate the performance of iotID in Section~\ref{sec:performance}, and discuss the related work in Section~\ref{sec:related-work}. We summarize the paper in Section~\ref{sec:summary}.


\section{Problem Statement and Data Collection} \label{sec:problemstatement}
In this section we first describe the problem of IoT device identification, and then we present the collection and pre-processing of the data in carrying out this study.

\subsection{Problem Statement}
We consider a network environment consisting of both IoT and non-IoT devices. This network environment can be a smart home, smart city, or any enterprise network with both kinds of devices deployed in the network. As discussed in Section~\ref{sec:introduction}, unlike non-IoT devices, there is a sheer diversity of manufacturers and models in IoT devices, and different IoT devices often perform very specific functionalities with different objectives. Given the diversity of IoT devices, they also require different levels of service guarantees and exhibit different levels of security properties. Consequently, network administrators would like to identify IoT devices in their networks for various performance or security purposes. For example, they may allow certain IoT devices to be used in the network but would like to block other IoT devices because of vulnerabilities associated with these IoT devices. 

An IoT device identification scheme such as iotID developed in this paper tries to identify the specific types of the IoT devices in the network, which is the critical first step in supporting device-specific policies adopted by the network. Schemes such as iotID can be deployed at the edge of a network by its administrator or its upstream network service provider.
As discussed above, a monitored network consists of both IoT and non-IoT devices; however, in this study we are only interested in identifying the specific types of IoT devices, given the unique challenges and opportunities in identifying IoT devices compared to the non-IoT devices. In this paper, we consider all the non-IoT devices as one single class, {\em non-IoT devices}; we do not identify the specific types of non-IoT devices.

\subsection{Data Collection and Pre-processing}
We collect the data used in this study in a typical smart home network, consisting of $12$ IoT devices and $7$ non-IoT devices. These devices are connected via a Wi-Fi router to the global Internet. The router is flashed with OpenWrt. In addition, tcpdump and some homemade scripts are installed on the router to capture network traffic on a daily basis, for $60$ continuous days. We randomly choose one week of data for this study (we have also performed studies using other weeks and similar results are obtained). The collected data contains both TCP and UDP flows. Each flow is identified by the standard 5 tuple of source and destination IP addresses, source and destination port numbers, and protocol, with a flow expiration threshold of $120$ seconds between two consecutive packets in the flow~\cite{mainuddin2021traffic}. Based on our preliminary studies, UDP flows only marginally improve the performance of iotID in classifying IoT devices, as a consequence, we only consider TCP flows in this study.

Table~\ref{tab:device-list} shows the types of IoT devices, their assigned type name used in this paper, and the number of TCP flows generated by the corresponding IoT devices. In the table, we also include one row for all $7$ non-IoT devices (including laptops, smartphones, and tablets). Since we are not interested in classifying the non-IoT devices, we consider them as a single type and named it as {\em non-IoT}.  We also note that we have two Logitech Circle-2 cameras in the network. As our focus is on identifying device types, both are considered as the same type and combined in the table. The same observation applies to the two Eufy Indoor cameras in the network. As a consequence, we have $12$ IoT devices on the network, but there are only $10$ different IoT device types. 

\begin{table*}[htbp]
\renewcommand{\arraystretch}{1.5}
\caption{Device Types and Traffic (One Week)}
\label{tab:device-list}
\centering
\begin{tabular}{|l|l|l|l|}
  \hline
  Device Type Name& Device Type & \# of TCP Flows & \% of Total TCP Flows \\
  \hline
  Non-IoT & All non-IoT devices & $253325$ & $90.5771$ \\
  \hline
  eufy\_cam & Eufy Indoor Cam 2K & $13369$ & $4.7801$ \\
  \hline
  amz\_echo & Amazon Echo Dot & $5399$ & $1.9304$ \\
  \hline
  wyze\_cam & Wyze Cam V2 & $4652$ & $1.6633$ \\
  \hline
  len\_spkr & Lenovo Smart Clock Speaker & $1132$ & $0.4047$ \\
  \hline
  hp\_printer & HP Officejet 3830 Printer & $1091$ & $0.3901$ \\
  \hline
  ltlelf\_cam & LittleElf Pan \& Tilt & $258$ & $0.0922$ \\
  \hline
  amz\_plug & Amazon Smart Plug & $254$ & $0.0908$ \\
  \hline
  semoic\_bulb & Semioc WiFi Smart Bulb & $88$ & $0.0315$ \\
  \hline
  logi\_cam & Logitech Circle 2 Camera & $76$ & $0.0272$ \\
  \hline
  epicka\_plug & Epicka Smart Plug & $35$ & $0.0125$ \\
  \hline
  \end{tabular}
\end{table*}

From the table we can see that devices generate drastically different amount of traffic in terms of the number of TCP flows (and also the number of packets and bytes, not shown in the table; see~\cite{mainuddin2021traffic} for more details on the network traffic characteristics of the devices). For example, while (the 12) IoT devices generate $26,354$ TCP flows, on average, IoT devices generate much less TCP flows compared to non-IoT devices ($253,325$ TCP flows, or about $90.58\%$ of all TCP flows). In addition, even among IoT devices, some IoT devices (such as Eufy camera) generate much more TCP flows than other IoT devices (such as Epicka smart plug). In essence, we can observe that the dataset is vastly imbalanced, with some device types containing much more traffic than other device types.



\section{Design of iotID} \label{sec:iotid}
In this section we will present the design of the iotID scheme and the features we extract from TCP flows to represent their traffic characteristics.

\subsection{iotID}
The scheme iotID is a machine-learning (ML) based IoT device identification system. At the high level, iotID will be trained based on a set of traffic features of known IoT devices, where each IoT device type will be considered as a separate class, and all non-IoT devices will be considered as another combined single class.  We use the device type name in Table~\ref{tab:device-list} as the class name in the ML model. The trained iotID scheme can then be deployed at the edge of a monitored network to identify the device types of other newly deployed IoT devices in the network. We will discuss the traffic features used in the development of iotID in the next subsection.

In our preliminary studies, we have experimented with a number of well-established machine learning algorithms, including SVM, Random Forest, xgboost, and MLP. Based on our preliminary studies, we note that in general xgboost outperforms all other ML algorithms that we have studied in most cases. Therefore, we will focus on xgboost in this study. Xgboost is an ensemble algorithm of decision trees (or more precisely CART) based on gradient boosting. Compared to other gradient boosting ensemble tree algorithms, xgboost adopted a number of algorithmic and system optimizations to improve its performance in terms of both resource consumption and classification accuracy~\cite{Chen:2016:xgboost}. 

In the design of iotID we need to handle the imbalance nature of the data traffic as discussed in the last section. There are a number of different methods to handle imbalanced datasets, including oversampling minority classes, undersampling majority classes, and cost-sensitive learning techniques, among others. Given that both oversampling and undersampling have some undesired drawbacks~\cite{HeMa2013:imbalancedlearning}, in this study we focus on the cost-sensitive learning techniques. More specifically, a cost or penalty is associated with a misclassification in cost-sensitive learning, which, instead of trying to optimize the overall accuracy, tries to minimize the overall misclassification cost. By associating a higher cost with a misclassification of a data point in a minority class than that of a majority class, a cost-sensitive learning algorithm will in essence put more efforts in minimizing misclassification of data points in minority classes. 

In this paper we adopt a simple {\em class balanced} cost mechanism, where our objective is that all classes will have the same cost or weight in optimizing the objective function of an ML algorithm. More specifically, let $N$ be the number of total data points in the dataset, $K$ the total number of classes in the dataset, $N\textsubscript{j}$ the number of data points in class $j$, $w\textsubscript{j}$ the weight assigned to a data point in class $j$, then $w\textsubscript{j}$ is given by:

    \begin{equation}\label{eqn-sample-weight}
        w_j = \frac{N}{KN_j}
    \end{equation}
We note that given this definition of $w_j$ to each data point in class $j$, the weight assigned to class $j$ is $w_j * N_j = N/K$, which is the same value for all classes, regardless of the number of data points in a class. Put in another way, all classes carry the same weight in optimizing the objective function of the ML algorithm. In addition to considering the imbalance nature of the dataset in the learning algorithm, we also consider this nature in the evaluation of the learning algorithm. We will discuss the performance metrics that we adopt to better illustrate the performance of iotID given the imbalanced dataset in the next section.
    
\addtolength{\topmargin}{0.04in}

\subsection{Feature Selection}
In this subsection we will discuss the features that we extract from each TCP flow in training and testing the ML algorithm of iotID. In order to effectively identify the IoT device type for a given TCP flow, we need to extract the features that can better represent or characterize TCP flows generated by various types of IoT devices (and non-IoT devices in a network with both types of devices). In our previous work~~\cite{mainuddin2021traffic} we have studied the network traffic characteristics of IoT (and non-IoT) devices from three complementary aspects: remote network servers and port numbers that IoT devices connect to, packet-level traffic characteristics such as packet inter-arrival time, and flow-level traffic characteristics such as flow duration, which provided critical insights into the operational and behavioral characteristics of IoT devices. 

Based on the insights obtained in~\cite{mainuddin2021traffic}, in this paper we will similarly extract traffic features of TCP flows from the three complementary aspects to represent a TCP flow. Overall we extract a total of $70$ features from each TCP flow.

\subsubsection{Remote network servers and port numbers ($3$ features)} Due to the autonomous nature of IoT devices, they normally only communicate with a small set of remote servers for uploading status data, accepting control commands, or updating software, among other operations. This is in drastic contrast with non-IoT devices, which may contact a larger and more diverse set of remote servers based on the browsing and other activities of device users. Therefore, we consider the IP address and port number of a TCP flow as important features. Given that it is common for an IoT device to contact multiple remote servers in the same network of the service provider (either a hosting service provider such as AWS or the manufacturer's own network), we split an IP address of a remote server into two parts (features): network prefix part and host part, based on BGP~\cite{rfc4271}. Consequently we have three features in this category.

\subsubsection{Packet level features ($54$ features)}
IoT (and non-IoT) devices exhibit different characteristics at the network packet level, in terms of packet sizes and inter-arrival times (IAT) of packets, due to the nature of the device firmware and autonomous behavior of IoT devices. However, different TCP flows (including the ones from the same device) may contain different number of packets. Based on our preliminary studies we note that the majority of TCP flows contain more than $10$ packets, therefore we choose the first $10$ packets in each direction of a TCP flow as features, including both packet sizes and IATs. If there are less than $10$ packets, we will fill the remaining ones with both sizes and IATs as zero. In this way we extract $38$ additional features from a TCP flow. In addition, we also extract a few statistical features including the minimum, maximum, mean, and standard deviation of the packet sizes and IATs in each direction of a TCP flow, which amount to $16$ additional features in this category.


\subsubsection{Flow level features ($13$ features)}
Different IoT devices generate very different amount of traffic in a TCP flow. In addition, while some IoT devices generate long-lasting TCP flows, others generate more bursty short TCP flows. We extract a number of features at the flow level that represent flow-level traffic characteristics of TCP flows. More specifically, for each TCP flow we extract the following features: flow duration; flow sizes in number of bytes in the incoming and outgoing directions of a flow, respectively; flow sizes in number of packets in the two directions, respectively; and similarly, flow rates in number of bytes and packets per second (in the two directions), respectively. Additionally, we also include the ratios of incoming and outgoing flow sizes (in bytes and packets) and rates (in bytes/sec and packets/sec), respectively. In this way we gather $13$ features at the TCP flow level.



\section{Performance Evaluation} \label{sec:performance}
\begin{figure*}[th]
        \begin{subfigure}[b]{0.32\textwidth}
                \includegraphics[width=\linewidth]{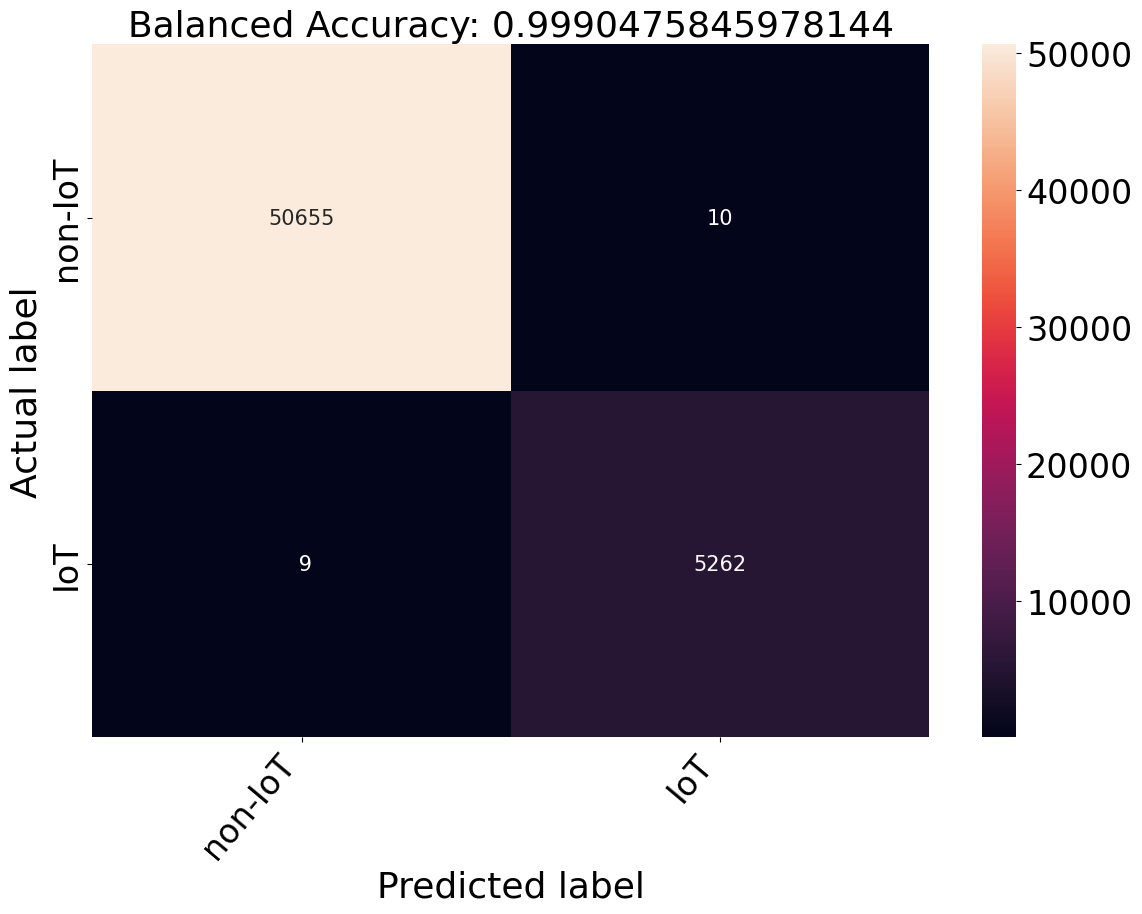}
                \caption{Stage-1 of Two-Stage iotID}
                \label{fig:stage-1-cm}
        \end{subfigure}%
        \hspace{\fill}
        \begin{subfigure}[b]{0.32\textwidth}
                \includegraphics[width=\linewidth]{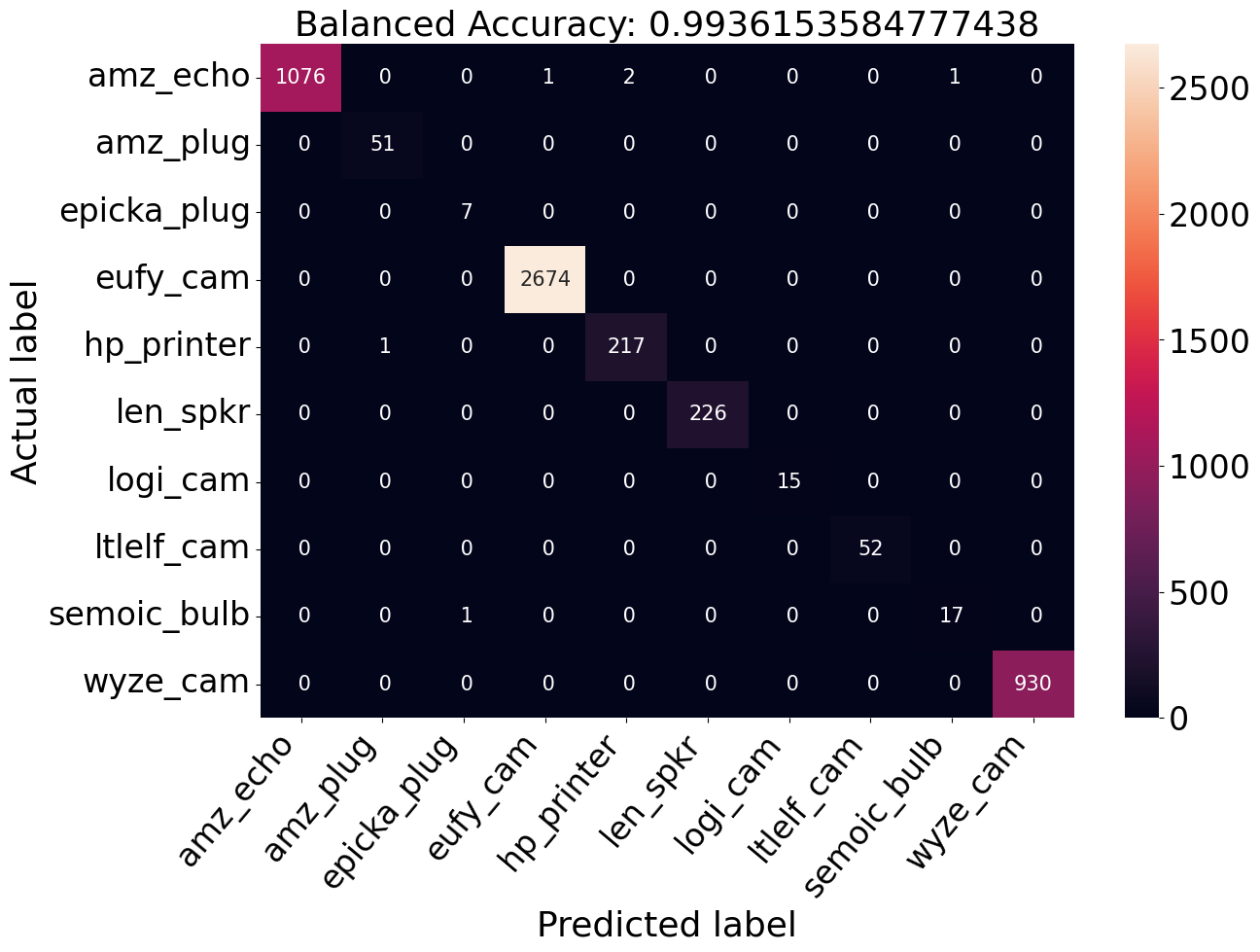}
                \caption{Stage-2 of Two-Stage iotID}
                \label{fig:stage-2-cm}
        \end{subfigure}%
        \hspace{\fill}
        \begin{subfigure}[b]{0.32\textwidth}
                \includegraphics[width=\linewidth]{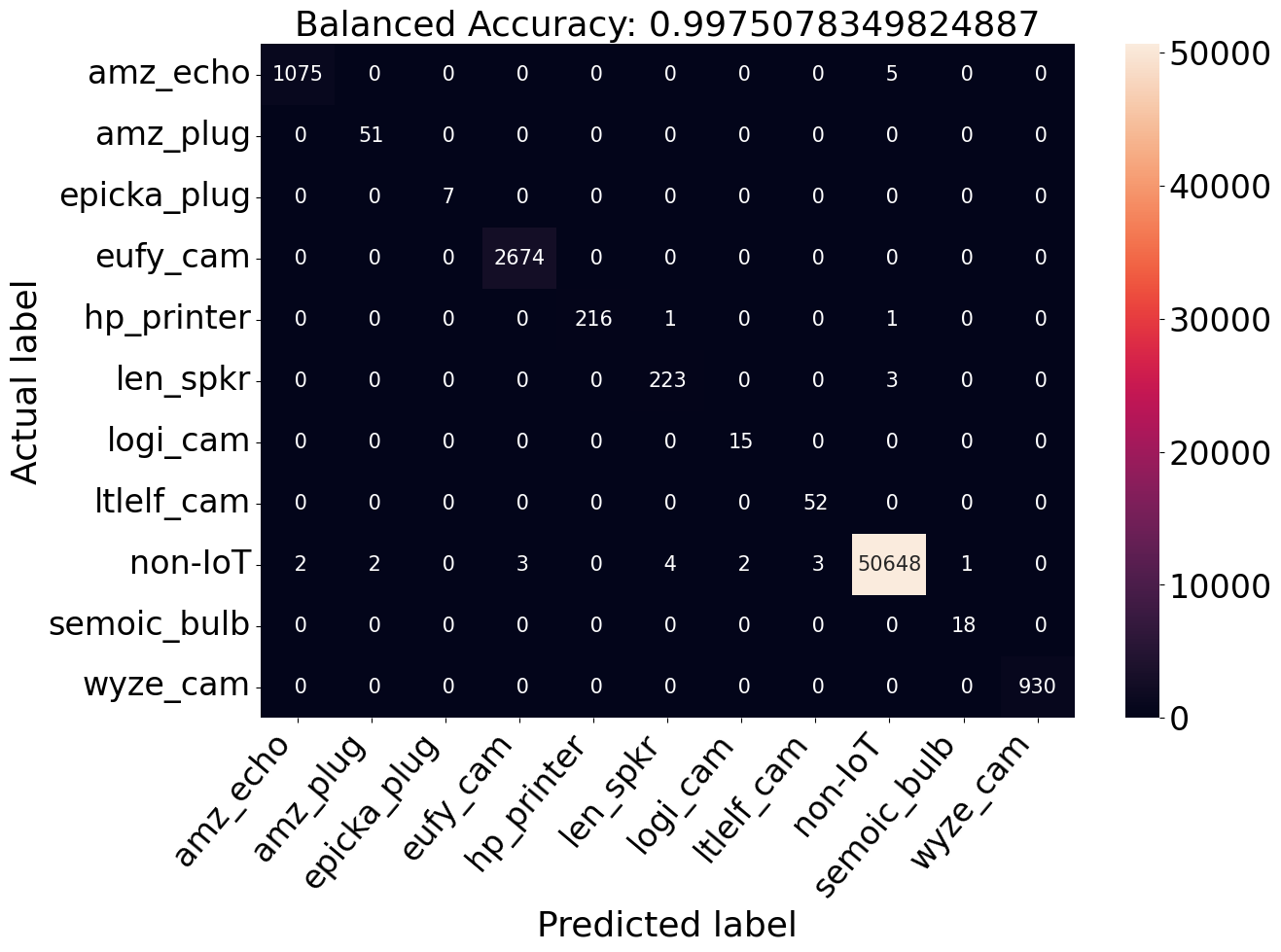}
                \caption{Single-Stage iotID}
                \label{fig:single-stage-cm}
        \end{subfigure}%
        \caption{Confusion Matrix.}\label{fig:confusion-matrix}
\end{figure*}
In this section we evaluate the performance of xgboost-based iotID using the dataset collected on the smart home network. We will first discuss the settings of the evaluation studies and then present the evaluation results.

\subsection{Evaluation Settings and Performance Metrics}
As we have discussed in Section~\ref{sec:problemstatement}, the dataset containing both IoT and non-IoT device traffic is imbalanced. 
In order to ensure that both training and test datasets contain the same proportion of data points from various device types as the original dataset, we adopt the stratified train-test split method to split the original dataset into a training set and a test set, containing $80\%$ and $20\%$ of the original data points, respectively. We use this single split of dataset to investigate the properties of iotID in terms of confusion matrix and feature importance (see below and the next subsection). In order to mitigate the potential bias in this single split of dataset, we also perform stratified 5-fold cross validation to better illustrate the performance of iotID. 

We consider two possible realizations of iotID in this paper. The first realization of iotID is a two-stage scheme. In the first stage, we aim to separate all IoT devices as a single class from non-IoT devices (recall that we consider all non-IoT devices as a combined single class). Put in another way, we treat this as a binary classification problem. In the second stage, we only focus on IoT devices and classify them into individual IoT device types. The second realization of iotID is a single-stage scheme, where each IoT device type is considered as a single class (and all non-IoT devices as a separate combined class). We will evaluate and compare the performance of the two realizations in the next subsection.

While the commonly used performance metric in machine learning is {\em accuracy}, which is defined as the proportion of correctly classified inputs, it is not an adequate metric for an imbalanced dataset~\cite{HeMa2013:imbalancedlearning}. In particular, the corresponding ML algorithm can bias towards majority classes if the objective is to simply optimize the accuracy. In order to better illustrate the performance of iotID given the imbalanced dataset, we adopt the {\em balanced accuracy} score (BAS), which is defined as the averaged score of recalls of individual classes in the dataset:

    \begin{equation}\label{eqn-bas}
        BAS = \frac{1}{K}\sum_{i = 1}^{K} \frac{TP_i}{TP_i + FN_i},
    \end{equation}
where $K$ is the total number of classes (device types), and $TP_i$ and $FN_i$ are the true positives and false negatives of class $i$, respectively. We note that in this definition the recalls of both majority classes and minority classes contribute in the same way in the balanced accuracy. In addition to the balanced accuracy, we also consider {\em confusion matrix} in order to better illustrate the performance of the scheme on individual classes.


\subsection{Two-Stage iotID}
As discussed above, in the two-stage realization of iotID, we first aim to separate all the IoT devices from non-IoT devices, and then in the second stage, we will only focus on identifying the specific types of IoT devices.

\begin{figure*}[th]
        \begin{subfigure}[b]{0.32\textwidth}
                \includegraphics[width=\linewidth]{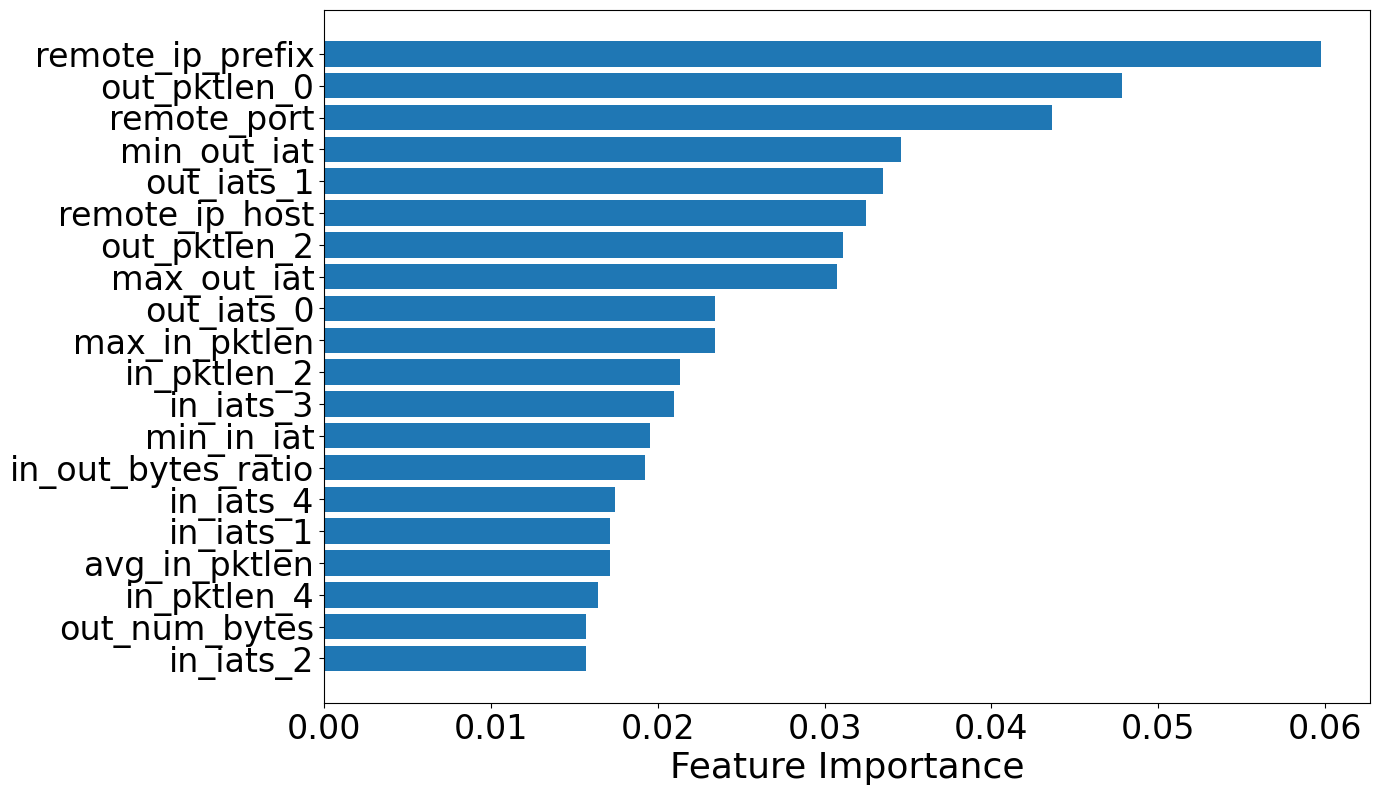}
                \caption{Stage-1 of Two-Stage iotID}
                \label{fig:stage-1-fi}
        \end{subfigure}%
        \hspace{\fill}
        \begin{subfigure}[b]{0.32\textwidth}
                \includegraphics[width=\linewidth]{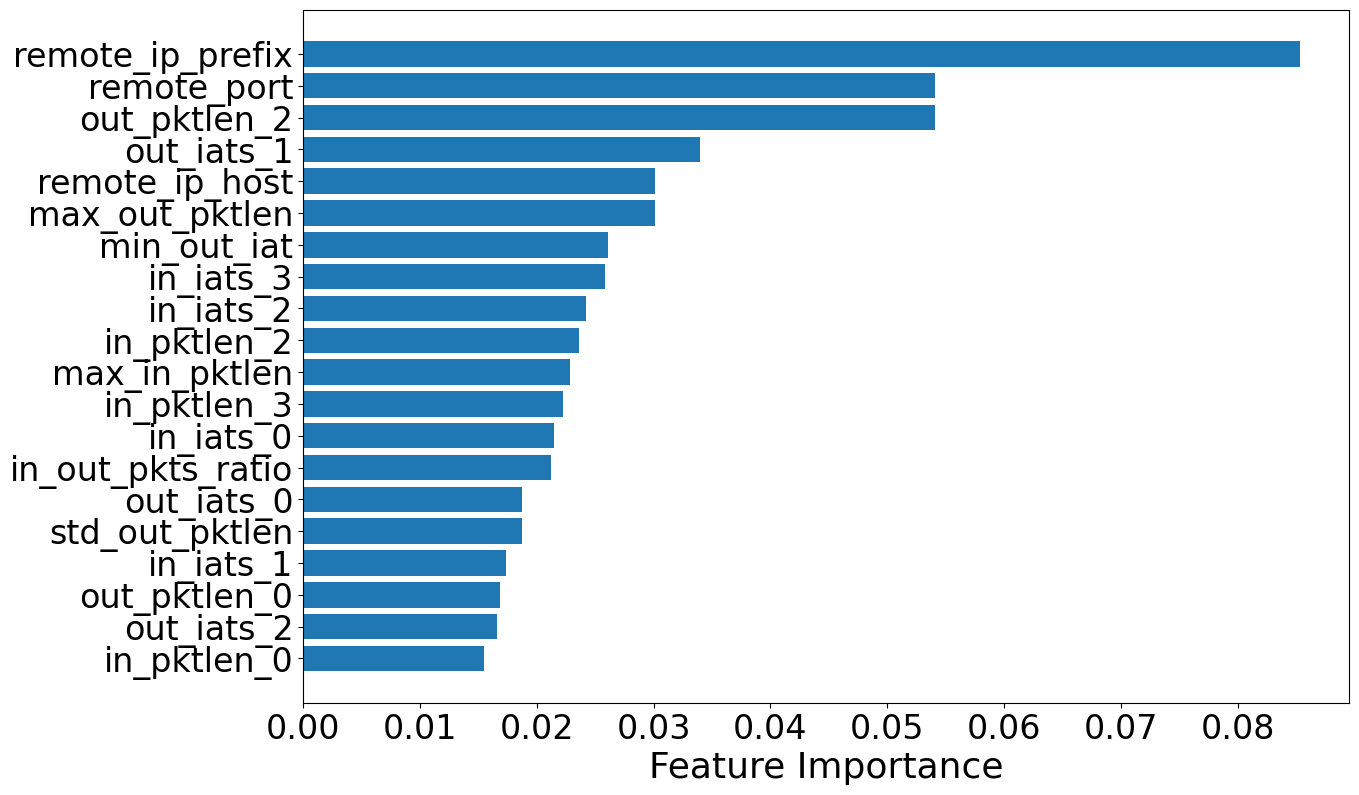}
                \caption{Stage-2 of Two-Stage iotID}
                \label{fig:stage-2-fi}
        \end{subfigure}%
        \hspace{\fill}
        \begin{subfigure}[b]{0.32\textwidth}
                \includegraphics[width=\linewidth]{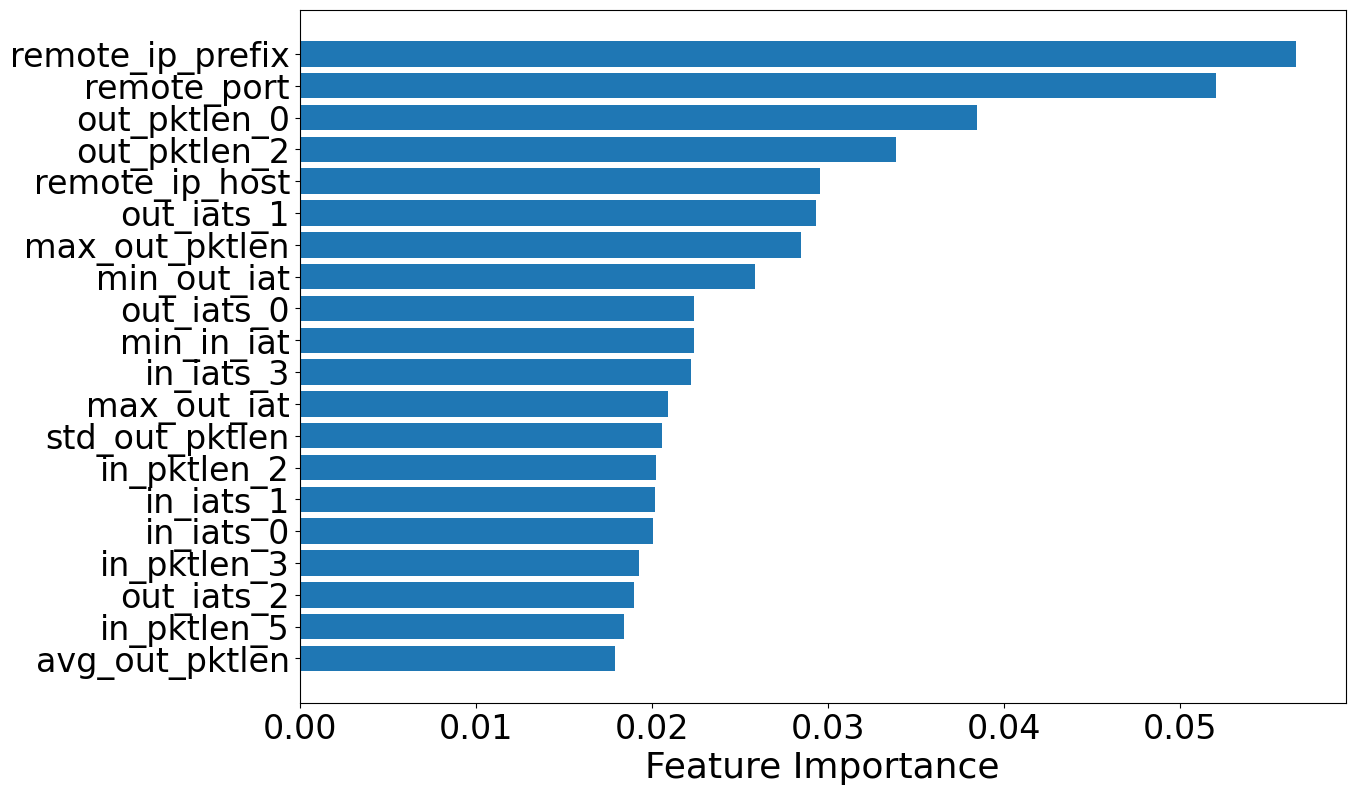}
                \caption{Single-Stage iotID}
                \label{fig:single-stage-fi}
        \end{subfigure}%
        \caption{Feature Importance.}\label{fig:feature-importance}
\end{figure*}


Figure~\ref{fig:stage-1-cm} shows the confusion matrix of the first stage of iotID. From the figure we can see that $9$ out of total $5271$ IoT TCP flows are misclassified, and $10$ out of $50665$ non-IoT TCP flows are misclassified, which results in a balanced accuracy of about $99.9\%$. Figure~\ref{fig:stage-1-fi} shows the significance of top features in affecting the performance of iotID. From the figure we can see that network prefix of remote server IP address (remote\_ip\_prefix) plays the most critical role in determining the iotID performance. This is understandable in that IoT devices normally only communicate with a small set of remote servers of service providers in order to carry out the specific functionalities of the devices~\cite{mainuddin2021traffic}. 

The figure also shows the contributions of other top features. In the name of the features, "out" means the packets in the outgoing direction of a TCP flow, and "in" means incoming direction. The numbers in a feature name indicates the index of a packet, with $0$ as the first packet. Others in the feature names are self-explanatory. Overall we can see that some other features such as remote port and properties of individual packets also play important role in determining the performance of stage-1 of iotID, consistent with observations made in~\cite{mainuddin2021traffic}.

Figure~\ref{fig:stage-2-cm} and~\ref{fig:stage-2-fi} shows the confusion matrix and feature importance of the second stage of the two-stage iotID, where we only handle data traffic of IoT devices. From the confusion matrix we can see that $4$ TCP flows from amz\_echo and $1$ flow from hp\_printer and semioc\_bulb are misclassified as other IoT types, which results in a balanced accuracy of about $99.4\%$. From the feature importance figure we can see again that remote network prefix plays the most important role in determining the performance of the second stage of the two-stage iotID, as they may use different service providers. 

Multiplying the performance of the two stages of iotID, we can see that the balanced accuracy of the two-stage realization of iotID is about $99.3\%$. We note that we have used all the TCP flows of IoT devices in the stage-2 of iotID evaluation, including misclassified TCP flows in the first stage, to maximize the data points in the second stage of iotID. This will only marginally affect the combined performance, given the small number of misclassified TCP flows of IoT devices in the first stage.

As discussed above, we also perform stratified 5-fold cross validation on the original dataset to investigate the performance of the two-stage iotID. The average balanced accuracy scores of the first stage and second stage of the two-stage iotID are $99.87\%$ and $99.45\%$, respectively. Combining the performance of the two stages, we can see that the average balanced accuracy of the two-stage iotID is about $99.3\%$, similar to the one we have observed using the single split of the original dataset without cross validation. In addition, we have also manually examined the balanced accuracy of each iteration in the 5-fold cross validation, and observed that they are very consistent (all are above $99\%$). These results indicate that the properties that we have observed in the single split of the dataset in terms of confusion matrix and feature importance should be representative of the two-stage iotID.

\subsection{Single-Stage iotID}
As discussed above, in the single-stage realization of iotID, we consider all non-IoT devices as one combined class, and each type of IoT devices also as a separate class. Figure~\ref{fig:single-stage-cm} shows the confusion matrix of the single-stage iotID. From the figure we can see that $17$ TCP flows of non-IoT devices are misclassified. We also note that $9$ TCP flows of IoT devices are misclassified as non-IoT flows (similar to stage-1 of the two-stage iotID), and an additional TCP flow of hp\_printer is misclassified as a flow of another IoT device (len\_spkr, a Lenovo speaker). The balanced accuracy of the single-stage iotID is about $99.8\%$, which is slightly better than the combined performance of the two-stage realization of iotID. 




Figure~\ref{fig:single-stage-fi} shows the importance of traffic features in determining the performance of the single-stage iotID. From the figure we can similarly see that features such as remote network prefix and port number play critical roles in affecting the performance of iotID. Similarly we have also studied the performance of the single-stage iotID with stratified 5-fold cross validation, and the average balanced accuracy is about $99.1\%$, slightly worse than what we have observed with the single split of the original dataset. Examining the individual balanced accuracy from each iteration of the 5-fold cross validation, we notice that the performance of one iteration (about $96.8\%$) is slightly worse than that of other iterations in the cross validation (above $99\%$). Overall, the confusion matrix and feature importance we have observed using the single split of the original dataset are still representative of the single-stage iotID, based on manual examination of the 5-fold cross validation results.

\section{Related Work}\label{sec:related-work}
In this section we will briefly discuss the research efforts on IoT device identification that are most relevant to our work. 

In~\cite{marchal2019audi}, the authors proposed a system to identify device type in an IoT network based on the periodic communication properties of the devices. Meidan {\it et al.} used device traffic properties such as network flow, protocols, and TCP sessions to classify IoT and non-IoT devices and to further identify the classes of different IoT device~\cite{meidan2017profiliot, meidan2017detection}. Miettinen {\it et al.} presented a study on IoT device fingerprinting and identification based on three categories of features, including first $N$ packets exchanged in the service discovery protocols~\cite{miettinen2017iot}. 


In~\cite{shahid2018recognition}, the authors presented an IoT device identification scheme based on network traffic behavior of the connected devices. Their feature set includes the packet sizes and inter-arrival-times of first $10$ packets from each TCP flow. Only $4$ IoT devices were used in their studies. In~\cite{aksoy2019automated}, the authors proposed a two-level classification approach of IoT devices, where Genetic Algorithm was used to select useful features among a large set of features extracted from a packet header, and then ML algorithms were used to classify the IoT device types. The authors of~\cite{chowdhury2020network} adopted a similar approach as above for device type identification using TCP/IP packet header information. Sivanathan {\it et al.} presented a machine-learning algorithm to classify IoT devices based on statistical attributes of their network activities~\cite{sivanathan2018classifying}, including features such as activity cycles, remote servers and ports, signaling patterns, and cipher suites. The authors of~\cite{sun2019automated} proposed a machine-learning based IoT device fingerprinting approach using general flow metadata and TLS handshake data. 

However, the majority of existing work only considered the classification of IoT devices without non-IoT devices. A real-world network normally contains both IoT and non-IoT devices, and it is critical to separate IoT devices from non-IoT devices in such an environment. In addition, none of existing work considered the imbalance nature of traffic generated by different IoT devices (and non-IoT devices) and their impacts on the performance of the IoT device identification algorithms. IotID takes into account the real-world deployment scenario with both IoT and non-IoT devices and the nature of imbalance traffic from various IoT and non-IoT devices.




\section{Summary and Future Work}\label{sec:summary}
In this paper we developed {\em iotID}, an effective machine-learning (ML) based IoT device identification scheme. In iotID, $70$ features of TCP flows from three complementary aspect were collected: remote network servers and port numbers, packet-level traffic characteristics such as packet inter-arrival time, and flow-level traffic characteristics such as flow duration. We evaluated the performance of iotID based on network traffic collected on a typical smart home environment consisting of both IoT and non-IoT devices, which showed that iotID can achieve a balanced accuracy score of above $99\%$. In our future work we will explore opportunities to evaluate iotID with additional IoT devices, and we will also thoroughly study the two realizations of iotID in various network settings to investigate the deployment scenarios where one may be preferred over another.


\bibliographystyle{unsrt}
\bibliography{iot_traffic}

\begin{thebibliography}{10}

\bibitem{mainuddin2021traffic}
Md~Mainuddin, Zhenhai Duan, and Yingfei Dong.
\newblock Network traffic characteristics of iot devices in smart homes.
\newblock In {\em 2021 International Conference on Computer Communications and
  Networks (ICCCN)}, pages 1--11, 2021.

\bibitem{Chen:2016:xgboost}
Tianqi Chen and Carlos Guestrin.
\newblock {XGBoost}: A scalable tree boosting system.
\newblock In {\em Proceedings of the 22nd ACM SIGKDD International Conference
  on Knowledge Discovery and Data Mining}, KDD '16, pages 785--794, New York,
  NY, USA, 2016. ACM.

\bibitem{HeMa2013:imbalancedlearning}
Haibo He and Yunqian Ma.
\newblock {\em Imbalanced Learning: Foundations, Algorithms, and Applications}.
\newblock Wiley-IEEE Press, 1st edition, 2013.

\bibitem{rfc4271}
Yakov Rekhter, Susan Hares, and Tony Li.
\newblock {A Border Gateway Protocol 4 (BGP-4)}.
\newblock RFC 4271, January 2006.

\bibitem{marchal2019audi}
Samuel Marchal, Markus Miettinen, Thien~Duc Nguyen, Ahmad-Reza Sadeghi, and
  N~Asokan.
\newblock Audi: Toward autonomous iot device-type identification using periodic
  communication.
\newblock {\em IEEE Journal on Selected Areas in Communications},
  37(6):1402--1412, 2019.

\bibitem{meidan2017profiliot}
Yair Meidan, Michael Bohadana, Asaf Shabtai, Juan~David Guarnizo, Mart{\'\i}n
  Ochoa, Nils~Ole Tippenhauer, and Yuval Elovici.
\newblock Profiliot: a machine learning approach for iot device identification
  based on network traffic analysis.
\newblock In {\em Proceedings of the symposium on applied computing}, pages
  506--509, 2017.

\bibitem{meidan2017detection}
Yair Meidan, Michael Bohadana, Asaf Shabtai, Martin Ochoa, Nils~Ole
  Tippenhauer, Juan~Davis Guarnizo, and Yuval Elovici.
\newblock Detection of unauthorized iot devices using machine learning
  techniques.
\newblock {\em arXiv preprint arXiv:1709.04647}, 2017.

\bibitem{miettinen2017iot}
Markus Miettinen, Samuel Marchal, Ibbad Hafeez, N~Asokan, Ahmad-Reza Sadeghi,
  and Sasu Tarkoma.
\newblock Iot sentinel: Automated device-type identification for security
  enforcement in iot.
\newblock In {\em 2017 IEEE 37th International Conference on Distributed
  Computing Systems (ICDCS)}, pages 2177--2184. IEEE, 2017.

\bibitem{shahid2018recognition}
M.~R. {Shahid}, G.~{Blanc}, Z.~{Zhang}, and H.~{Debar}.
\newblock Iot devices recognition through network traffic analysis.
\newblock In {\em 2018 IEEE International Conference on Big Data (Big Data)},
  pages 5187--5192, 2018.

\bibitem{aksoy2019automated}
Ahmet Aksoy and Mehmet~Hadi Gunes.
\newblock Automated iot device identification using network traffic.
\newblock In {\em IEEE International Conference on Communications (ICC)}, pages
  1--7. IEEE, 2019.

\bibitem{chowdhury2020network}
Rajarshi~Roy Chowdhury, Sandhya Aneja, Nagender Aneja, and Emeroylariffion
  Abas.
\newblock Network traffic analysis based iot device identification.
\newblock In {\em Proceedings of 4th International Conference on Big Data and
  Internet of Things}, pages 79--89, 2020.

\bibitem{sivanathan2018classifying}
Arunan Sivanathan, Hassan~Habibi Gharakheili, Franco Loi, Adam Radford, Chamith
  Wijenayake, Arun Vishwanath, and Vijay Sivaraman.
\newblock Classifying iot devices in smart environments using network traffic
  characteristics.
\newblock {\em IEEE Transactions on Mobile Computing}, 18(8):1745--1759, 2018.

\bibitem{sun2019automated}
Jianhua Sun, Kun Sun, and Chris Shenefiel.
\newblock Automated iot device fingerprinting through encrypted stream
  classification.
\newblock In {\em International Conference on Security and Privacy in
  Communication Systems}, pages 147--167. Springer, 2019.

\end{thebibliography}

\vspace{10pt}

\end{document}